\documentclass[aps,prb,superscriptaddress,twocolumn,nofootinbib]{revtex4-2}
\usepackage[utf8]{inputenc}
\usepackage{textgreek}
\usepackage{amsmath, amssymb}
\usepackage{graphicx}
\usepackage{dcolumn}
\usepackage{bm}
\usepackage{xcolor}
\usepackage{hyperref}
\usepackage{braket}
\usepackage[section]{placeins}
\usepackage{physics}
\usepackage[T1]{fontenc}
\usepackage[strings]{underscore}

\begin{document}

\title{Quantum simulation of molecular excited-state manifolds and energies using the TEPID-ADAPT-VQE algorithm}

\author{Jason Saroni}
\email{jsaroni@vt.edu}
\affiliation{Department of Physics, Virginia Tech, Blacksburg, VA 24061, USA}
\affiliation{Virginia Tech Center for Quantum Information Science and Engineering, Blacksburg, VA 24061, USA}

\author{Bharath Sambasivam}
\affiliation{Department of Physics, Virginia Tech, Blacksburg, VA 24061, USA}
\affiliation{Virginia Tech Center for Quantum Information Science and Engineering, Blacksburg, VA 24061, USA}

\author{Ayush Asthana}
\affiliation{Department of Chemistry, University of North Dakota, Grand Forks, ND 58201, USA}

\author{Edwin Barnes}
\affiliation{Department of Physics, Virginia Tech, Blacksburg, VA 24061, USA}
\affiliation{Virginia Tech Center for Quantum Information Science and Engineering, Blacksburg, VA 24061, USA}

\date{\today}

\begin{abstract}
The simulation of molecular excited states is a key challenge in quantum chemistry and a promising application for quantum computing. In this work, we investigate the efficacy of the truncated eigenvalue parametrized initial density adaptive variational algorithm (TEPID–ADAPT-VQE) for computing low-lying excited states and potential energy surfaces. TEPID–ADAPT variationally diagonalizes a truncated low-temperature Gibbs state, enabling the simultaneous preparation of multiple excited states within a single optimization. We apply the method to H$_2$, LiH, and linear H$_4$ across bond lengths spanning weakly and strongly correlated regimes. The adaptive derivative-assembled problem-tailored (ADAPT) ansatz construction yields compact circuits suitable for near-term hardware. We also implement a modified version of the algorithm MORE-ADAPT-VQE \cite{Grimsley_2025} for comparison with TEPID-ADAPT. We find that both algorithms accurately reproduce excited-state spectra and potential energy curves within chemical accuracy for all the molecules and geometries studied. However, TEPID-ADAPT has the advantage of utilizing only a single, physically motivated hyperparameter (temperature) that controls the energy scale at which excited states are targeted, while MORE-ADAPT utilizes multiple hyperparameters whose optimal values depend sensitively on the target problem. These results demonstrate that combining adaptive ansatz construction with density-matrix-based formulations provides an efficient framework for excited-state quantum chemistry on near-term devices.
\end{abstract}

\maketitle

\section{Introduction}

Electronic excited states play a central role in molecular processes driven by light and high-energy interactions, governing phenomena such as photochemistry, spectroscopy, and nonadiabatic dynamics~\cite{Armentrout1991,Knepp2025,Jankowska2021,liu2020mapping,curchod2024perspective}. Accurate access to excited-state spectra is therefore essential both for predictive chemical modeling and for testing the limits of electronic structure theory. However, obtaining reliable and predictively accurate excitation energies and their potential energy surfaces remains challenging, particularly in regimes where the ground state has a strong multi-configurational character~\cite{gonzalez2012progress,herbert2024visualizing}.

Traditional classical approaches, including variants of multi-reference methods, such as the Complete Active-Space Self-Consistent Field (CASSCF)~\cite{olsen2011casscf, evangelista2018perspective}, are limited in the size of the active space that can be modelled using a classical computer; while scalable methods based on linear response, such as EOM-CCSD and TD-DFT offer limited accuracy across the entire potential energy surface, when the excited state is dominated by double excitations, or when the ground state is strongly correlated~\cite{Yordanov2022,Gao2024,Knowles1989,LiLu2020, rishi2017excited,laurent2013td,lischka2018multireference}. While these methods remain standard approaches currently used, the computational cost of the highly accurate and reliable approaches, such as CASSCF with large active spaces, rapidly becomes prohibitive for strongly correlated molecules or for systematic exploration of excited-state manifolds. This limitation has motivated the development of alternative approaches capable of capturing excited-state physics using quantum computers.

Recent advances in hardware have spurred significant interest in quantum algorithms for electronic structure problems. Variational quantum algorithms, in particular, offer a promising framework for near-term quantum simulation due to their flexibility and relatively shallow circuit requirements~\cite{Higgott2019,PhysRevResearch.1.033062,Asthana2023}. Extending these methods to excited states has been an active area of research, with approaches ranging from subspace-search and folded-spectrum techniques to equation-of-motion and thermal-state-based formulations~\cite{Gocho2023,CadiTazi2024,Dutta2026,Asthana2023,kumar2023quantum,ollitrault2020quantum,mcclean2020decoding,urbanek2020chemistry}.

A key challenge for excited-state algorithms lies in balancing ans\"atz expressibility with quantum resource constraints. Fixed-form ans\"atze, such as standard unitary coupled cluster constructions, often struggle to describe strongly correlated excited states without incurring prohibitive circuit depth~\cite{Yordanov2021}. Adaptive ansatz construction provides a principled solution to this challenge by systematically selecting operators that are most relevant to the target problem. The adaptive derivative-assembled problem-tailored (ADAPT) framework has been shown to produce compact and accurate circuits for ground-state calculations~\cite{Grimsley2019}, and recent work, such as the MORE-ADAPT algorithm, has demonstrated its potential for excited-state simulations as well~\cite{GrimsleyEvangelista2025}.

In this work, we apply a recent ADAPT-based algorithm developed originally to prepare the Gibbs state, the truncated eigenvalue parametrized initial density (TEPID–ADAPT) method, to compute molecular excited state potential energy surfaces. The TEPID-ADAPT algorithm aims to determine multiple low-lying excited states within a single variational optimization loop. It constructs a unitary that maps a rank-truncated density matrix in the computational basis to an approximate Gibbs state. The same unitary also simultaneously maps the computational basis states to the low-lying eigenstates of the problem~\cite{Sambasivam2025}.

We benchmark TEPID-ADAPT on molecular systems exhibiting a range of correlation regimes, including bond stretching and near-degeneracy. Our analysis focuses on both the spectral completeness and the accuracy of each potential energy surface. By integrating ADAPT-based approaches into excited-state formulations, we demonstrate that chemically accurate excited-state energies can be obtained with compact circuits suitable for near-term quantum hardware. We also compare the performance of TEPID-ADAPT against that of an alternative variational algorithm for preparing excited states called MORE-ADAPT-VQE~\cite{Grimsley_2025}. We consider a version of MORE-ADAPT in which the weights that determine the relative importance of different contributions to the cost function are allowed to differ from one another. We find that although MORE-ADAPT does equally well in preparing the excited states of the molecules and geometries we consider, the performance is sensitive to the weights, and it is not clear a priori how to choose them. In contrast, TEPID-ADAPT utilizes only a single hyperparameter---the temperature---whose physical meaning and algorithmic impact are clear---it determines the energy scale at which to search for excited states.

The remainder of this paper is organized as follows. Sec.\ref{sec:methodology} introduces the molecular Hamiltonians and algorithmic frameworks employed in this study. Benchmarking results for representative molecular systems are presented in Sec.~\ref{sec:results}. Finally, Sec.~\ref{sec:Discussion_and_outlook} summarizes the main findings and outlines the directions for future work.

\FloatBarrier

\section{Methodology}
\label{sec:methodology}

The electronic structure Hamiltonians considered in this work are expressed in second quantization as
\begin{equation}
\hat{H}=\sum_{ij}h_{ij}\hat{c}_{i}^{\dagger}\hat{c}_{j}+\frac{1}{2}\sum_{ijkl}h_{ijkl}\hat{c}_{i}^{\dagger}\hat{c}_{j}^{\dagger}\hat{c}_{k}\hat{c}_{l},
\label{eq:Ham}
\end{equation}
where the fermionic creation and annihilation operators $\hat{c}^{\dagger}_i$ and $\hat{c}_i$ create and annihilate an electron in spin-orbital $i$, respectively. The coefficients $h_{ij}$ and $h_{ijkl}$ denote the one- and two-electron integrals, which depend on the molecular geometry and encode the underlying electronic interactions.

To simulate the Hamiltonian on a quantum computer, the fermionic operators must be mapped to qubit operators that can be implemented using quantum gates. In this work, we employ the Jordan--Wigner transformation \cite{Jordan1928}, noting that alternative mappings are also available \cite{BRAVYI2002210}. Under the Jordan--Wigner mapping, the fermionic operators are represented as
\begin{equation}
\begin{aligned}
\hat{c}_{j}^{\dagger} &= \frac{1}{2}\left(\hat{X}_{j}-i\hat{Y}_{j}\right)\prod_{k<j}\hat{Z}_{k}, \\
\hat{c}_{j} &= \frac{1}{2}\left(\hat{X}_{j}+i\hat{Y}_{j}\right)\prod_{k<j}\hat{Z}_{k},
\end{aligned}
\label{eq: JW}
\end{equation}
where $\hat{X}_i$, $\hat{Y}_i$, and $\hat{Z}_i$ are Pauli operators acting on the qubit associated with spin-orbital $i$. Applying this mapping transforms the Hamiltonian in Eq.~\eqref{eq:Ham} into a sum of Pauli strings,
\begin{equation}
\hat{H}=\sum_{l}\eta_{l}\hat{P}_{l},
\label{eq: Ham2}
\end{equation}
where $\hat{P}_{l}\in\left\{ \hat{I},\hat{X},\hat{Y},\hat{Z}\right\}^{\bigotimes N}$ are nontrivial Pauli strings acting on $N$ qubits, and $\eta_l$ are the corresponding coefficients. The second-quantized Hamiltonians in Eq.~\eqref{eq:Ham} are generated using the \texttt{Qiskit Nature} open-source framework \cite{qiskit_nature_2023}. We now describe the excited-state algorithms employed in this study.

\subsection{TEPID-ADAPT}

The TEPID-ADAPT algorithm seeks a unitary transformation that diagonalizes a truncated density matrix and maps computational basis states to approximate eigenstates of the Hamiltonian. The target density matrix is approximated as
\begin{align}
\rho_G \approx \frac{1}{Z_m}\sum_{k=1}^{m}e^{-\beta E_k} \ket{\psi_k}\bra{\psi_k},
\end{align}
where $Z_m$ denotes the truncated partition function up to state $m$, $\ket{\psi_k}$ is the $k$-th eigenstate, and $E_k$ is the corresponding eigenvalue \cite{Sambasivam2025}. The adaptive ansatz is constructed to variationally diagonalize this density matrix within the truncated subspace. The cost function is the free energy 

\begin{align}
F(\vec{\mu}, \vec{\theta}) =  \left< H \right>(\vec{\mu}, \vec{\theta}) - \beta^{-1}\mathcal{S}(\vec{\mu}),
\end{align}
where $\vec{\theta}$ are the adaptive variational parameters, $\beta$ is the inverse temperature, $\mu_k$ are variational parameters that approximate $\frac{e^{-\beta E_k}}{Z_m}$ at convergence with $\sum_{k=1}^m\mu_k=1$, and $\mathcal{S}$ is the von Neumann entropy. The energy expectation value with respect to the density matrix $\left< H \right>$ is the only part of the cost function that is measured on the quantum computer \cite{Sambasivam2025}---the entropy does not need to be measured, as it can be calculated classically from knowledge of the $\mu_k$. In the context of excited states, the inverse temperature, $\beta$ serves as a physically motivated hyperparameter in the algorithm that sets the energy scale over which to search for eigenstates.

\begin{figure*}[ht!]                
\includegraphics[width=\linewidth]{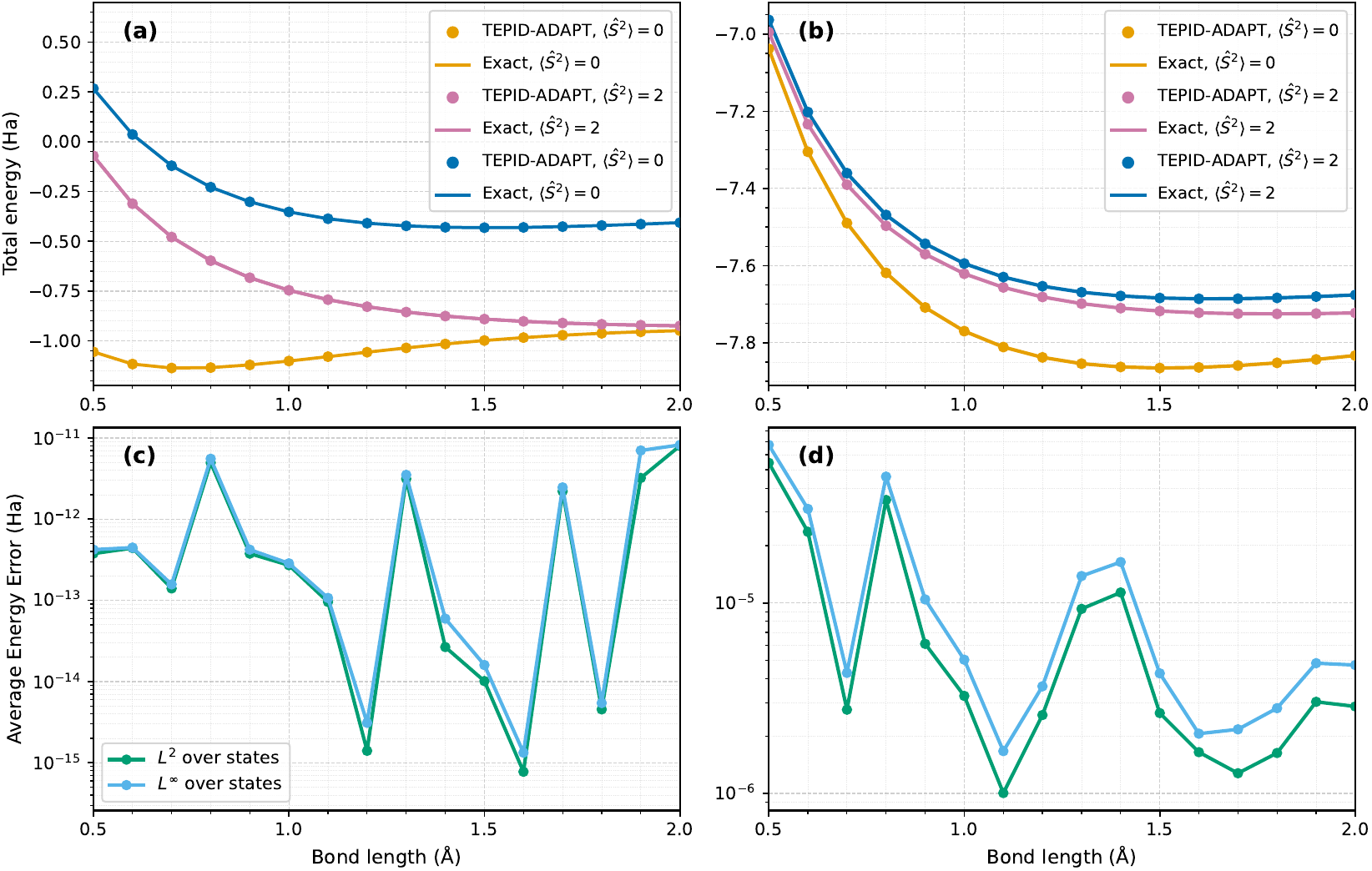}
\caption{(a) Energies of the lowest excited states of H$_2$ obtained from TEPID-ADAPT with an inverse temperature of $\beta=0.5$ and from exact diagonalization. All but the pink curve have $\langle\hat{S}^2\rangle=0$. (b) The potential energy curve for LiH where the orange line depicts the ground state energy and the pink and blue lines the lowest excited-state energies with $\langle\hat{S}^2\rangle=2$. In panels (c) and (d), the average errors defined in Eq.~\eqref{eq:linfty_err} and Eq.~\eqref{eq:l2_err} are shown as functions of bond length for TEPID-ADAPT at $\beta=0.5$ and are within $10^{-4}$ Ha. The closeness of $L^2$ and $L^{\infty}$ errors suggests that the errors are nearly uniform across the states.}
\label{fig:H2_total_energies_error_summary}
\end{figure*}

\subsection{Weighted MORE-ADAPT}

The MORE-ADAPT-VQE algorithm employs the cost function \cite{Grimsley_2025,PhysRevResearch.1.033062}
\begin{align}
C(\vec{\theta}) = \sum_{k=1}^m w_k
\left \langle c_k\left| U_S^{\dagger}(\vec{\theta}) \hat{H} U_S(\vec{\theta}) \right |c_k \right \rangle,
\end{align}
where the weights $w_k$ determine the relative importance of the target eigenstates, $m$ is the number of eigenstates targeted, and $\ket{c_k}$ denotes a chosen reference state. Note that all weights must be chosen up front for MORE-ADAPT---in contrast to a single temperature control variable for TEPID-ADAPT---and how to determine optimal choices for these weights remains an open question \cite{Hong2023RefiningTW, Ding2024groundexcitedstates}. For each problem we consider, we choose the weights from the range $[0,1]$ such that $w_i>w_j$ when $i<j$ \cite{PhysRevResearch.1.033062}. This is a slight generalization of the original MORE-ADAPT algorithm, where the weights were chosen uniformly: $w_k=1/m$. The unitary ansatz $\hat{U}_S(\vec{\theta}) = \prod_{n=1}^{N_{\theta}} e^{\theta_n\hat{A}_n}$ is constructed adaptively by iteratively selecting operators $\hat{A}_n$ from a predefined pool based on the magnitude of their associated energy gradients. Here $N_{\theta}$ denotes the total number of selected pool operators and $\theta_n$ is purely imaginary.

The gradients used to guide operator selection are evaluated for each
candidate operator $\hat A_{\mu}$ as
\begin{align}
\left.
\frac{\partial C_{\mu}(\theta_{\mu})}{\partial \theta_{\mu}}
\right|_{\theta_{\mu}=0}
=
\sum_{k=1}^m w_k
\left\langle c_k \right|
U_S^{\dagger}(\vec{\theta})
\left[\hat H,\hat A_{\mu}\right]
U_S(\vec{\theta})
\left| c_k \right\rangle,
\end{align}
where
\begin{align}
C_{\mu}(\theta_{\mu})
=
\sum_{k=1}^m w_k
\left\langle c_k \right|
U_S^{\dagger}(\vec{\theta})
e^{-\theta_{\mu}\hat A_{\mu}}
\hat H
e^{\theta_{\mu}\hat A_{\mu}}
U_S(\vec{\theta})
\left| c_k \right\rangle.
\end{align}
At each iteration, the operator with the largest gradient magnitude is appended to the ansatz, with its corresponding variational parameter initialized to zero. Following each operator addition, all variational parameters are re-optimized to minimize the cost function. The Broyden-Fletcher-Goldfarb-Shannon (BFGS) algorithm is used in all parameter optimization routines \cite{Fletcher1987PMO}.

\subsection{Error evaluation}

To quantify the accuracy of the computed excited-state energies, we define the absolute error with respect to exact energies for the $k$-th state at internuclear distance $d$ as
\begin{align}
\varepsilon_k(d) = \left|E_k^{\text{exact}}(d)-E_k^{\text{algorithm}}(d)\right|,
\end{align}
where the algorithm is either weighted MORE-ADAPT or TEPID-ADAPT. From this, we define the $L^{\infty}$ error
\begin{align}
\varepsilon_{L^{\infty}}(d)
= \max\limits_{i=1,\ldots,m} \varepsilon_i(d),
\label{eq:linfty_err}
\end{align}
and the root-mean-square ($L^2$) error
\begin{align}
\varepsilon_{L^{2}}(d) = \sqrt{\frac{1}{m}\sum_{k=1}^{m}\varepsilon_k(d)^2}.
\label{eq:l2_err}
\end{align}
Here, $\varepsilon_{L^{\infty}}$ corresponds to the $L^{\infty}$ norm and captures the worst-case deviation among all states, while $\varepsilon_{L^2}$ corresponds to the $L^2$ norm and provides a measure of the typical error across the excited-state manifold at a given bond length.

\section{Results}
\label{sec:results}

We benchmark the TEPID-ADAPT and weighted MORE-ADAPT algorithms on molecular hydrogen (H$_2$), LiH, and linear H$_4$ chains across internuclear separations. These systems provide a controlled setting in which excited-state structure and spin symmetry play a central role, while still allowing exact diagonalization for quantitative comparison. All reported energies are compared against exact eigenvalues obtained via exact diagonalization of the qubit Hamiltonian.

\subsection{\texorpdfstring{Excited-state potential energy curves of $\mathrm{H}_2$ and $\mathrm{LiH}$}{Excited-state potential energy curves of H2 and LiH}}

Figure~\ref{fig:H2_total_energies_error_summary}(a) presents the lowest physically relevant excited-state energy levels of the H$_2$ molecule as functions of internuclear separation using a minimal basis of STO-3G (Slater-Type Orbital – 3 Gaussian) \cite{Hehre1969}, as obtained using TEPID-ADAPT. Here, we measure the expectation values of the total spin-squared operator,
$\langle \hat{S}^2 \rangle$, and the $z$-component of the total spin operator,
$\langle \hat{S}_z \rangle$. For simultaneous eigenstates of
$\hat{S}^2$ and $\hat{S}_z$, these expectation values are equal to
$S(S+1)$ and $S_z$, respectively, where $S$ and $S_z$ denote the total spin
and spin projection quantum numbers. We set $\hbar=1$ throughout. With the exception of a single triplet state ($\langle\hat{S}^2\rangle = 2$), all states shown correspond to singlet symmetry ($\langle \hat{S}^2 \rangle = 0$). As the bond is stretched, we observe a near-degeneracy, which is characteristic of bond dissociation and the emergence of strong correlation~\cite{10.1063/1.1682533}. We construct low-depth quantum-computer-constructible Hartree--Fock-inspired reference states with well-defined $S_z$ that are mapped to eigenstates by TEPID-ADAPT. For example, taking H$_2$ from \texttt{Qiskit Nature} with little-endian notation and an ordering with spin-up electrons on the first half and spin-down electrons on the second half of the state, $\ket{0101}$, $\ket{0011}$, $\ket{1001}$, and $\ket{1100}$ have $\langle \hat{S}_z \rangle$ values of $0$, $-1$, $0$, and $1$, respectively, that map to the $\langle \hat{S}^2 \rangle=0$ ground state and $\langle \hat{S}^2\rangle=2$ first excited states. The energies that correspond to these states and the next lowest excited-state energy with $\langle \hat{S}^2 \rangle=0$ are shown in Fig.~\ref{fig:H2_total_energies_error_summary}(a). Similarly for LiH, the Hartree--Fock state $\ket{0001100011}$ maps to the ground state and the $\langle \hat{S}_z\rangle=0$ state $\ket{0001100101}$ to the first excited state with $\langle\hat{S}^2\rangle=2$. The next lowest excited state energy as a function of distance obtained from the reference state $\ket{0010100101}$ with $\langle \hat{S}^2\rangle=2$ and $\langle\hat{S}_z\rangle=0$ is shown in Fig.~\ref{fig:H2_total_energies_error_summary}(b). In both (a) and (b), the $S_z$-preserving qubit-excitation-based (QEB) pool \cite{Yordanov2021} is used. This pool was proposed as an alternative to fermionic excitation pools in adaptive VQE methods and includes the following operators:
\begin{align}
\hat{A}_{ik}
=
\frac{1}{2}
\left(
\hat{X}_i \hat{Y}_k - \hat{Y}_i \hat{X}_k
\right)
\end{align}

\begin{align}
\hat{A}_{ijkl}
&=
\frac{1}{8}\Big(
\hat{X}_i \hat{Y}_j \hat{X}_k \hat{X}_l
+ \hat{Y}_i \hat{X}_j \hat{X}_k \hat{X}_l
\notag \\
&\quad
+ \hat{Y}_i \hat{Y}_j \hat{Y}_k \hat{X}_l
+ \hat{Y}_i \hat{Y}_j \hat{X}_k \hat{Y}_l
- \hat{X}_i \hat{X}_j \hat{Y}_k \hat{X}_l
\notag \\
&\quad
- \hat{X}_i \hat{X}_j \hat{X}_k \hat{Y}_l
- \hat{Y}_i \hat{X}_j \hat{Y}_k \hat{Y}_l
- \hat{X}_i \hat{Y}_j \hat{Y}_k \hat{Y}_l
\Big)
\end{align}
corresponding to single and double qubit-excitation operators, where $i$, $j$, $k$, and $l$ index the qubits acted upon by the operator.

The TEPID-ADAPT algorithm accurately reproduces the exact excited-state eigenenergies across the entire range of bond lengths studied. In particular, the convergence of the orange and pink curves in Fig.~\ref{fig:H2_total_energies_error_summary}(a), at large separations for H$_2$, reflects the correct dissociation limit where electronic configurations become energetically degenerate. Notably, the algorithm remains numerically stable even at extended bond lengths for LiH and H$_2$, where standard variational approaches often suffer from convergence difficulties due to near-degenerate manifolds. Appendix~\ref{sec:conv_plots} contains plots showing how the energy error decays as the ansätze grow for H$_2$ and LiH both at bond distance $d = 0.40$~\AA. 

Figures~\ref{fig:H2_total_energies_error_summary}(c) and (d) quantify corresponding errors at each distance through the $L^2$ and $L^{\infty}$ deviations from exact energies. Across all bond lengths, both error metrics remain well within $10^{-4}$ Ha. The $L^2$ error consistently tracks below the $L^{\infty}$ error, indicating that deviations are not dominated by outlier states but are uniformly small across the excited-state manifold. Importantly, no systematic degradation in accuracy is observed as the bond length increases, demonstrating that the adaptive ansatz construction effectively captures both weakly and strongly correlated regimes.

These results highlight a key advantage of the TEPID-ADAPT approach: by optimizing a single cost function that incorporates multiple reference states, the method robustly resolves near-degeneracies and maintains accuracy without requiring state-by-state optimization.

\begin{figure}[htp]                
\includegraphics[width=\linewidth]{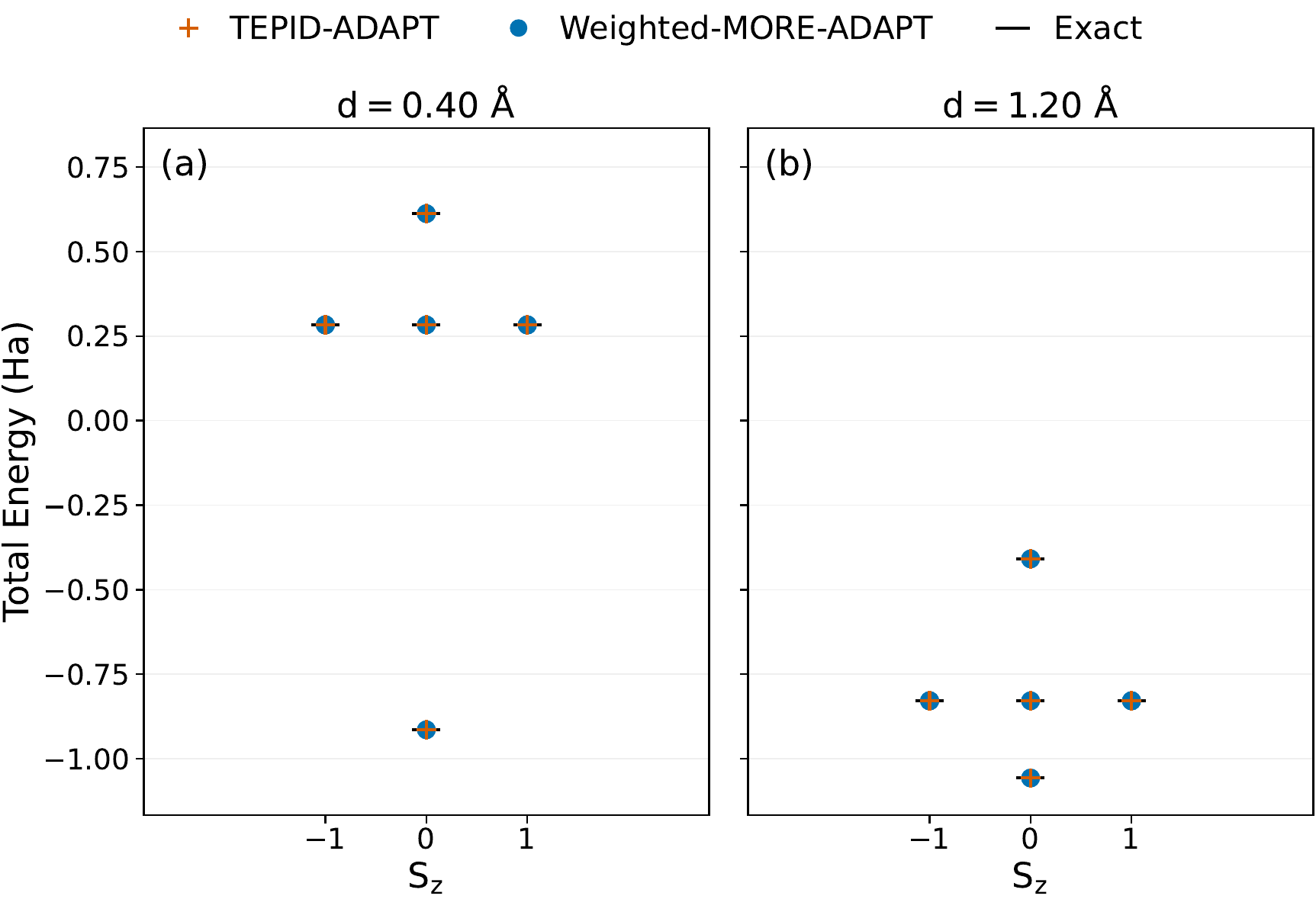}
\caption{The energy spectrum of H$_2$ at bond distances of (a) 0.4~\AA{} and (b) 1.2~\AA{} obtained from TEPID-ADAPT, weighted MORE-ADAPT, and exact diagonalization. The reference states for TEPID-ADAPT and weighted MORE-ADAPT are chosen as computational basis states, and the full Pauli pool is used for the adaptive procedure. Plots of the energy error for the TEPID-ADAPT results can be found in Appendix~\ref{sec:appendix_errors}. The weights $w_i > w_j$ for $i < j$, with $w_k \in [0.5,1]$, have been used. The weights are defined as $w_k = 1 - \frac{k}{30}$ for $k = 0, 1, \ldots, 15$, yielding the sequence $1.0000, 0.9667, 0.9333, \ldots, 0.5333, 0.5000$ to four significant digits. Only states satisfying physical constraints on $S$, $S_z$, and particle number are included in the plots. The inverse temperature $\beta=0.5$ was used for TEPID-ADAPT. The spin multiplicities are determined by evaluating $\langle\hat{S}^2\rangle$.}
\label{fig:H2_energy_level_structure1}
\end{figure}

\subsection{Excited-state spectra of linear H chain}

Figures~\ref{fig:H2_energy_level_structure1} and~\ref{fig:energy_level_structure2} compare excited-state energies of linear H$_2$ and H$_4$ chains at two representative geometries: a compressed configuration ($d = 0.40$~\AA) and a moderately stretched configuration ($d = 1.20$~\AA) using the minimal STO-3G basis. Exact eigenvalues are shown as horizontal reference lines, while numerical results from TEPID-ADAPT and weighted MORE-ADAPT are overlaid. We choose the weights $w_k = 1 - \frac{k}{30}$ for $k = 0, 1, \ldots, 15$, giving the sequence $1.0000, 0.9667, 0.9333, \ldots, 0.5333, 0.5000$ to four significant digits.

From Fig.~\ref{fig:energy_level_structure2} at the compressed geometry, the spectrum exhibits well-separated excited states with distinct spin quantum numbers. Both algorithms accurately map computational basis states onto the corresponding eigenstates, reproducing the correct energy ordering and spin assignments. States belonging to the same symmetry sector cluster tightly around the exact energies. At the stretched geometry, the spectrum becomes significantly denser due to increased correlation and the proliferation of near-degenerate states. Despite this increased complexity, both weighted MORE-ADAPT-VQE and TEPID-ADAPT continue to recover the correct eigenenergies with high accuracy. This demonstrates that both methods can handle truncated but highly structured excited-state manifolds without loss of accuracy.

The close agreement of the two algorithms with exact results across both geometries indicates that adaptive variational algorithms are effective at preparing excited states. By selecting operators based on instantaneous gradients, both methods efficiently tailor the unitary to the relevant subspace, while preserving spectral resolution.

\subsection{Average energy accuracy}
\subsubsection{Temperature dependence}
Figure~\ref{fig:H4_d0p4_error_summary_vs_beta} examines the accuracy of excited states obtained via TEPID-ADAPT as a function of inverse temperature $\beta$ for H$_4$. The $L^2$ and $L^{\infty}$ energy errors are shown. As $\beta$ increases, corresponding to lower effective temperatures, both the $L^2$ and $L^{\infty}$ errors increase. Errors in higher excited states grow because a circuit trained on a low-temperature Gibbs state captures only low-lying states accurately. This behavior reflects the increased sensitivity of low-temperature thermal states to inaccuracies in the diagonalizing unitary for higher excited states. This is due to the stronger Boltzmann suppression of these higher excited states. At large $\beta$, the thermal distribution becomes sharply peaked around the lowest-energy states, amplifying the impact of any residual off-diagonal contributions in the approximate density matrix. For $\beta$ values relevant to chemically meaningful excited-state populations, TEPID-ADAPT maintains energy accuracy. 

\begin{figure}[htp]                
\includegraphics[width=\linewidth]{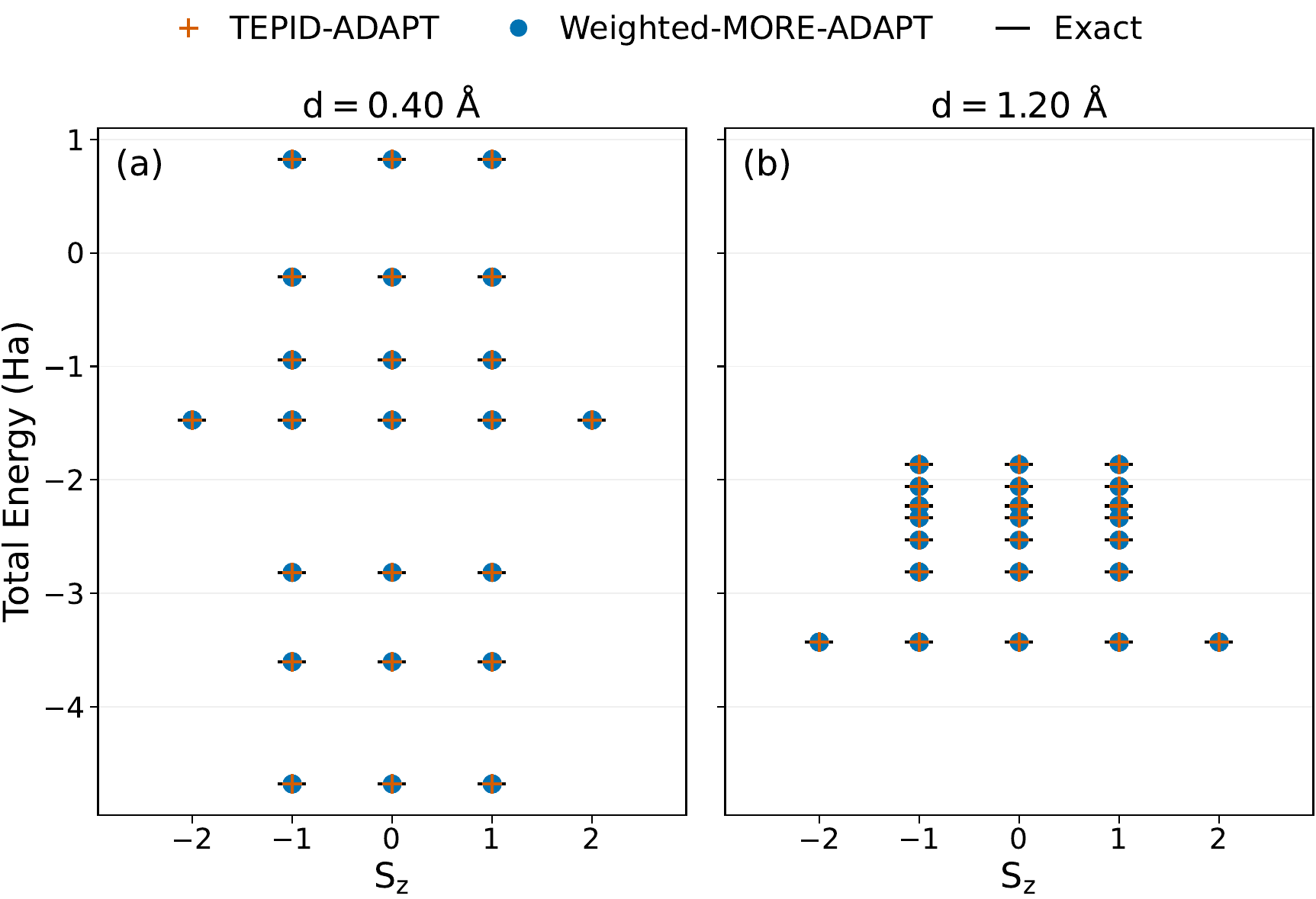}
\caption{Excited-state energies from TEPID-ADAPT and weighted MORE-ADAPT-VQE are compared with results from exact diagonalization. (a) Results for a compressed internuclear distance between H$_4$ atoms of $d=0.40$~\AA{}, and (b) results for distance $d=1.20$~\AA{} are shown. Computational basis states are used as reference states for both TEPID-ADAPT and weighted MORE-ADAPT, and both algorithms employ the full Pauli pool. Again the spin multiplicities are determined by evaluating $\langle\hat{S}^2\rangle$. The weights are defined as $w_k=1-\frac{k}{30}$, $k=0, 1, ..., 15$, yielding the sequence $1.0000, 0.9667, 0.9333, \ldots, 0.5333, 0.5000$ to four significant digits and the plots include only states with physically admissible values of $S$, $S_z$, and particle number. Plots of the energy error for the TEPID-ADAPT results can be found in Appendix B.}
\label{fig:energy_level_structure2}
\end{figure}

\subsubsection{Weight choice}

With weighted MORE-ADAPT, the absolute error of the lowest-indexed energy level increases on average by a factor of approximately $5 \times 10^{4}$ when transitioning from weights $w_i > w_j$ for $i < j$, with $w_k \in [0.5,1]$ to equal weights $w_k=\frac{1}{m}$ for the H$_4$ system at $d = 1.2$~\AA{}.

\begin{figure}[t]                
\includegraphics[width=\linewidth]{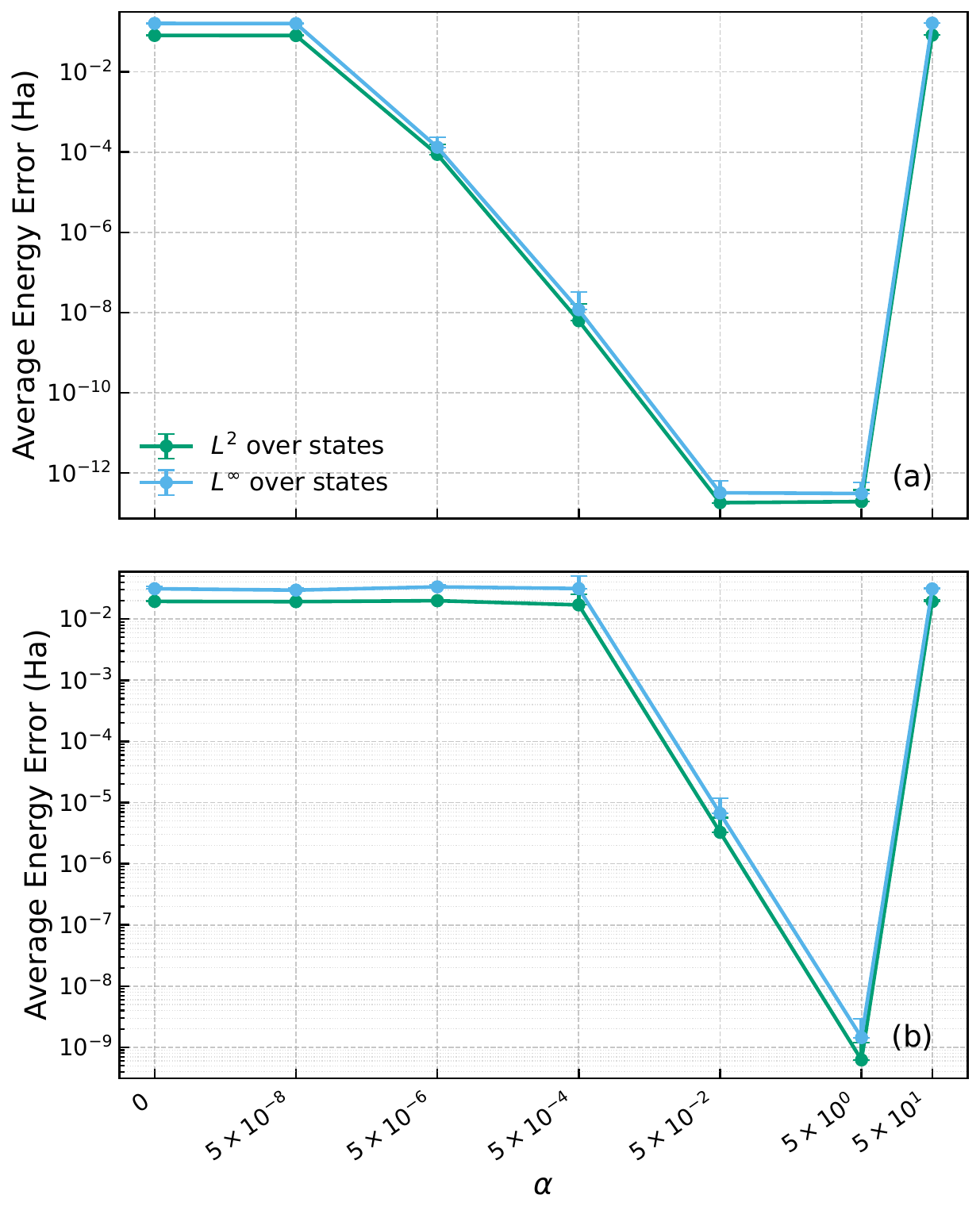}
\caption{Average energy error as a function of the parameter $\alpha$ for (a) $\mathrm{H}_2$ and (b) $\mathrm{H}_4$ obtained from weighted MORE-ADAPT.
Both the $L^2$ and $L^\infty$ error metrics are shown on a logarithmic scale. The normalized weights $w_k$ have been selected so that $w_k \propto e^{-\alpha \lambda_k}$ where $\lambda_k$ correspond to Hartree--Fock determinant energy values. For H$_2$, the Hartree--Fock determinant-energy vector is
$\vec{\lambda} = [-2.2273,-1.4821,\ldots,0.8730]$,
whereas for H$_4$ it is
$\vec{\lambda} = [-3.9148,-3.7696,\ldots,-3.0539]$. The plotted results for H$_2$ at $d = 0.4$~\AA{} and H$_4$ at $d = 1.2$~\AA{} are over 3 and 7 distinct physical states respectively from weighted MORE-ADAPT. Error bars denote the standard deviation obtained across 10 runs.}
\label{fig:errors_vs_alpha}
\end{figure}

Figure~\ref{fig:errors_vs_alpha} shows results from weighted MORE-ADAPT for the average energy error for H$_2$ and H$_4$ as a function of a parameter $\alpha$ that controls the distribution of weights via $w_k=\frac{e^{-\alpha\lambda_k}}{N}$. We use $\vec{\lambda}=[\lambda_1,\lambda_2,\ldots,\lambda_m]$ to denote a Hartree--Fock determinant-energy vector, where $\lambda_k$ is the Hartree--Fock energy of the $k$th determinant. Here, we choose $\vec{\lambda}=[-2.2273,-1.4821,\ldots,0.8730]$ for H$_2$, $\vec{\lambda}=[-3.9148,-3.7696,\ldots,-3.0539]$ for H$_4$, and $N$ is a normalization constant chosen such that $\sum_{k=1}^m w_k=1$. The precise $\alpha$ for which specified accuracy is reached differs depending on the system studied. From Figs.~\ref{fig:errors_vs_alpha}(a) and (b), while decreasing $\alpha$ improves accuracy in both systems, $\mathrm{H}_2$ requires smaller $\alpha$ to start reaching lower errors, whereas $\mathrm{H}_4$ starts achieving better accuracy at larger $\alpha$. This disparity demonstrates that selecting $\alpha$ may require problem-specific tuning.

\section{Discussion and Outlook}
\label{sec:Discussion_and_outlook}
The results presented in this work demonstrate that ADAPT-based excited-state algorithms provide a compelling balance between numerical accuracy and quantum resources. Across all benchmark systems studied, both TEPID-ADAPT and weighted MORE-ADAPT reliably reproduce excited-state spectra within chemical accuracy, conventionally defined as approximately 1~kcal/mol ($\sim$1.6~mHa)~\cite{Pople1999,Dobrautz2023}. Achieving this level of accuracy is particularly significant in the context of excited-state calculations, where near-degeneracies and strong correlation effects pose substantial challenges for both classical and quantum methods.

\begin{figure}[t]                
\includegraphics[width=\linewidth]{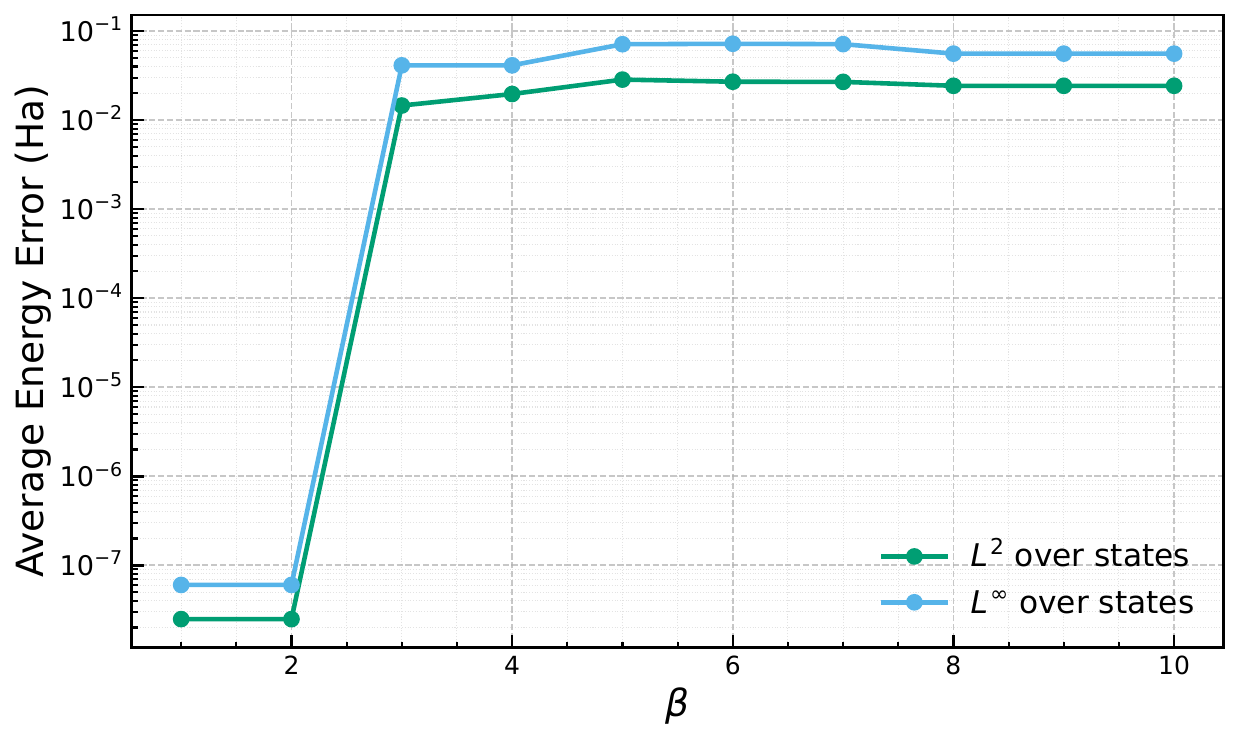}
\caption{The $L^2$ and $L^{\infty}$ errors as functions of the inverse temperature $\beta$ for simulations of TEPID-ADAPT applied to the problem of preparing 23 eigenstates of H$_4$ at a bond distance of 0.4~\AA, for which the corresponding energies are shown in the left panel of Fig.~\ref{fig:energy_level_structure2}}
\label{fig:H4_d0p4_error_summary_vs_beta}
\end{figure}

%
%

For molecular hydrogen and LiH, TEPID-ADAPT accurately resolves excited-state potential energy curves over a wide range of internuclear separations, including regions exhibiting near-degeneracies. Such features are well known to accompany bond dissociation and the emergence of strong (static) correlation, as the electronic wavefunction acquires pronounced multiconfigurational character~\cite{TruhlarPES,Fuchs2005,HollettGill2011}. The ability of a single adaptive ansatz to remain stable and accurate across these regimes highlights a key advantage of the TEPID-ADAPT formulation that includes optimization over entropy: multiple excited states are optimized simultaneously, mitigating state-tracking difficulties commonly encountered in traditional state-by-state variational approaches.

For the linear H$_4$ system, both TEPID-ADAPT and weighted MORE-ADAPT successfully recover truncated excited-state manifolds. Despite the increased density of states and enhanced correlation at larger interatomic separations, the adaptive construction of the diagonalizing unitary enables accurate mapping between computational basis states and physical eigenstates. The close agreement between the two algorithms across distinct correlation regimes suggests that the adaptive selection of operators plays a more central role than the specific excited-state cost function in determining overall performance. The thermal-state-based TEPID-ADAPT results further elucidate the trade-offs inherent in finite-temperature and mixed-state approaches. As the inverse temperature $\beta$ increases, corresponding to lower effective temperatures, both the $L^2$ and $L^{\infty}$ energy errors increase. This behavior reflects the growing sensitivity of low-temperature Gibbs states to residual off-diagonal errors in the approximate diagonalizing unitary~\cite{Sambasivam2025}.

Nevertheless, for moderate $\beta$ values relevant to chemically meaningful excited-state populations, TEPID-ADAPT maintains energy accuracy while avoiding ancilla qubits and excessive circuit depth. These features make it a practical candidate for near-term implementations on NISQ hardware. While weighted MORE-ADAPT is equally effective at finding excited-state energies for all the target problems and geometries we considered, achieving this performance required setting the weights appropriately, and at present, there is no known generally applicable systematic method for doing this. We saw that the performance of MORE-ADAPT is sensitive to this choice, which could become problematic in applying this method to larger problems. In contrast, TEPID-ADAPT's single hyperparameter, the temperature, can be viewed intuitively as setting the energy scale at which we want to search for excited states, thus avoiding the ambiguities present in MORE-ADAPT.

More broadly, the results reinforce the importance of adaptive ansatz construction for excited-state quantum algorithms. Fixed-form ans\"atze, such as conventional UCC variants, often struggle to balance expressibility and circuit depth, particularly in strongly correlated regimes~\cite{Yordanov2021,Anastasiou2024}. In contrast, ADAPT-based methods systematically tailor the ansatz to the problem at hand, yielding compact circuits that preserve essential physics while minimizing resource overhead. This advantage is especially pronounced for excited states, where capturing near-degenerate subspaces is essential.

Looking forward, several avenues merit further investigation. Extending these methods to larger active spaces and chemically relevant polyatomic molecules will be crucial for assessing scalability. Incorporating symmetry-adapted pools---such as the unitary coupled-cluster generalized singles and doubles pool \cite{Lee2019}---and hardware-efficient operator pools similar to the QEB pool we utilize here may further reduce circuit depth and improve robustness to noise~\cite{Tang2021,Ramoa2025}. 
Additionally, integrating error mitigation techniques and hardware-aware transpilation strategies will be necessary to translate the demonstrated algorithmic performance to experimental quantum devices.

In summary, this work provides quantitative evidence that ADAPT-based excited-state algorithms, namely TEPID-ADAPT and weighted MORE-ADAPT can achieve highly accurate potential energy surfaces for challenging molecular systems.
By unifying adaptive ansatz construction with both subspace-based and thermal-state formulations, these methods offer a flexible, scalable, and black-box style framework for excited-state quantum chemistry on near-term quantum hardware.



\begin{acknowledgments}
This work was supported by the National Science Foundation, award no. 2427046. EB also acknowledges support by the U.S. Department of Energy, Office of Science, Office of Advanced Scientific Computing Research under Award Number DE-SC0025430.
\end{acknowledgments}

\setlength{\textfloatsep}{10pt}

\bibliography{refs}

\onecolumngrid

\begin{appendix}
\section{Convergence Plots}
\label{sec:conv_plots}
\begin{figure*}[ht!]    
\includegraphics[width=\linewidth]{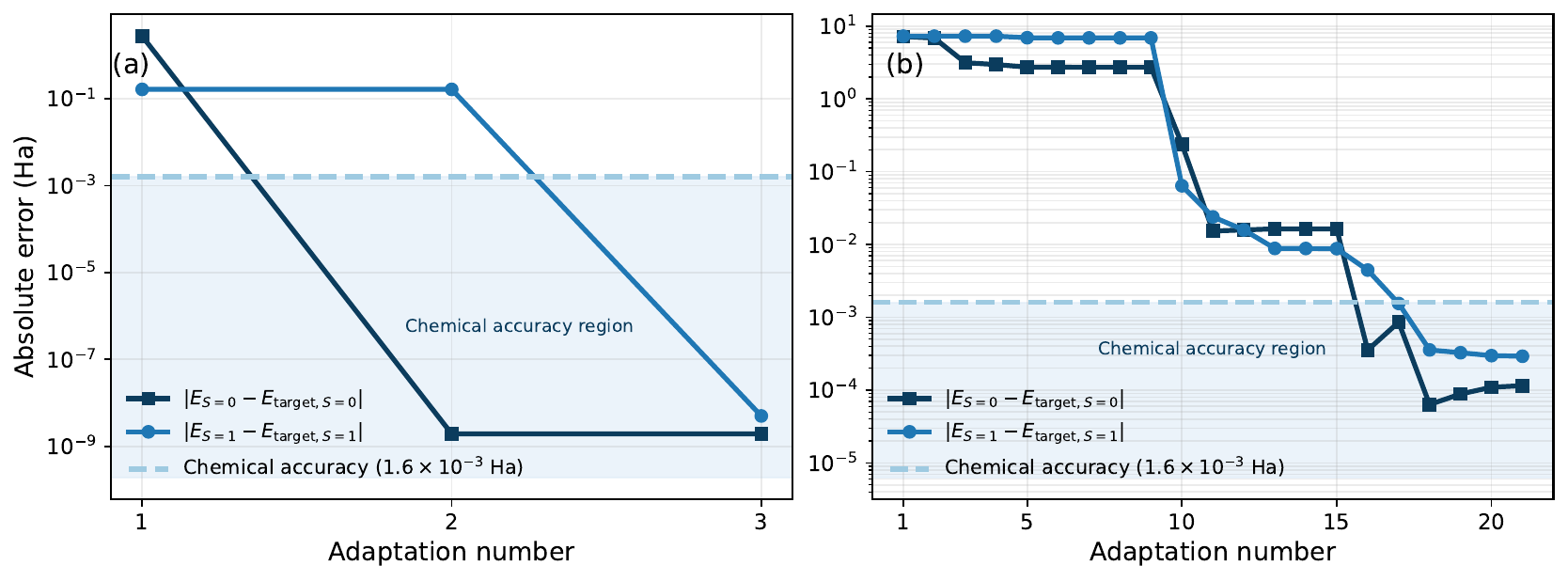}
\caption{Energy error versus adaptation number for TEPID-ADAPT applied to (a) H$_2$ at bond distance $d = 0.4$~\AA{} and (b) LiH at the same bond distance. In both cases, we start with $\langle\hat{S}_z\rangle=0$ states and use the $S_z$-preserving QEB pool. In (a) initial states $\ket{0101}$ and $\ket{1001}$ are used while in (b) $\ket{0001100011}$ and $\ket{0010100101}$ are used.}
\label{fig:combined_conv}
\end{figure*}

Figures~\ref{fig:combined_conv}(a) and (b) show how energy errors decay as TEPID-ADAPT iteratively grows ansätze for H2 and LiH excited-state manifolds. We start with $\langle\hat{S}_z\rangle=0$ states and use the $S_z$-preserving QEB pool. In Fig.~\ref{fig:combined_conv}(a) initial states $\ket{0101}$ and $\ket{1001}$ are used, while in Fig.~\ref{fig:combined_conv}(b) $\ket{0001100011}$ and $\ket{0010100101}$ are used. In Fig.~\ref{fig:combined_conv}(b) around the middle of the adaptation sequence, the TEPID-ADAPT algorithm identifies operators that strongly couple the reference states to the relevant low-lying energy correlated subspace. This leads to a rapid drop in the energy error by several orders of magnitude. Ultimately convergence is seen in both Figs.~\ref{fig:combined_conv}(a) and (b). 

\section{Absolute Energy Errors}
\label{sec:appendix_errors}

Here we present the absolute energy errors associated with the computed energy states for H$_2$ and H$_4$, providing a quantitative assessment of the accuracy of the methods discussed in the main text.

\begin{figure*}[ht!]    
\includegraphics[scale=0.45]{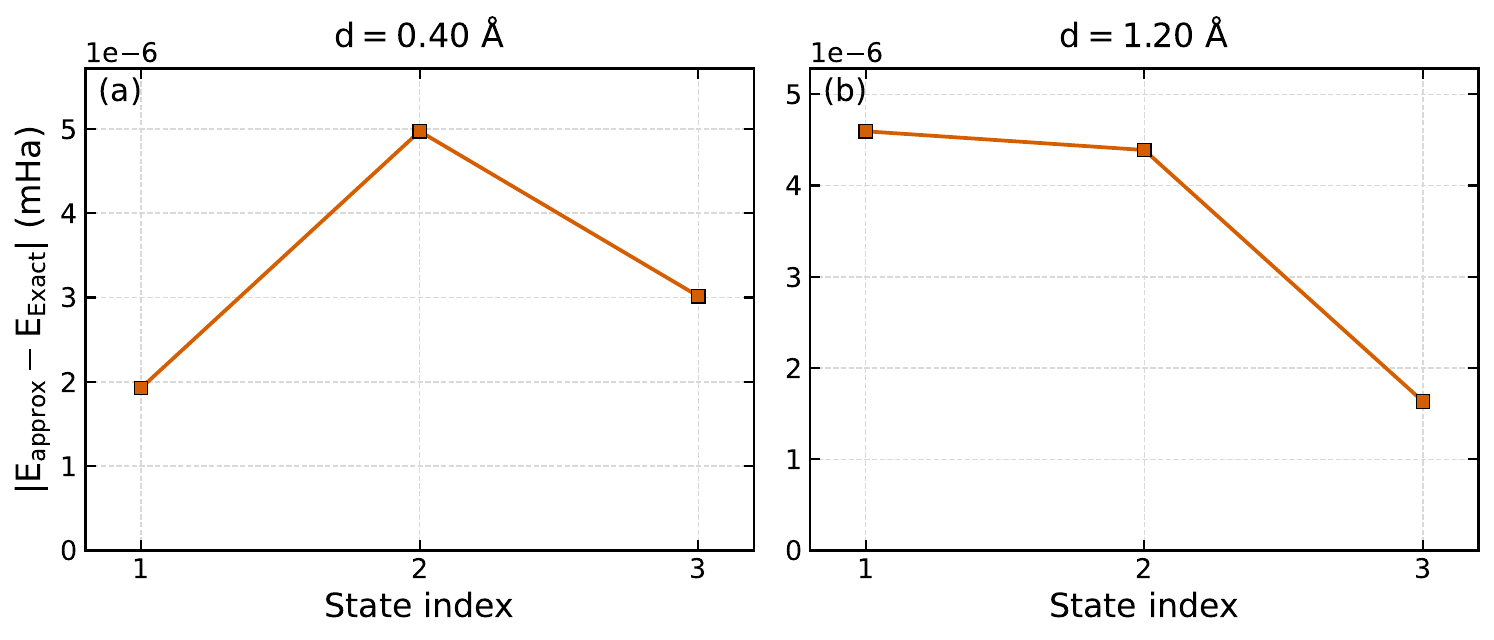}
\caption{Absolute energy errors for the energy eigenstates of H$_2$ corresponding to Fig.~\ref{fig:H2_energy_level_structure1}. Errors are shown in mHa as a function of state index. The graphs show consistently small deviations across states, indicating high overall proximity to exact results within chemical accuracy at (a) $d=0.40$~\AA{} and (b) $d=1.20$~\AA{}.}
\label{fig:H2_error}
\end{figure*}

Figure~\ref{fig:H2_error} shows the absolute energy errors for the energy eigenstates of H$_2$, corresponding to Fig.~\ref{fig:H2_energy_level_structure1}. The errors are reported in milli-Hartree (mHa) as a function of the state index. Across all states, the deviations from the exact energies remain consistently small. In particular, the results demonstrate high overall accuracy, with errors remaining within chemical accuracy at both $d=0.40$~\AA{} and $d=1.20$~\AA{} by the TEPID-ADAPT algorithm.

\begin{figure*}[ht!]   
\centering
\includegraphics[scale=0.45]{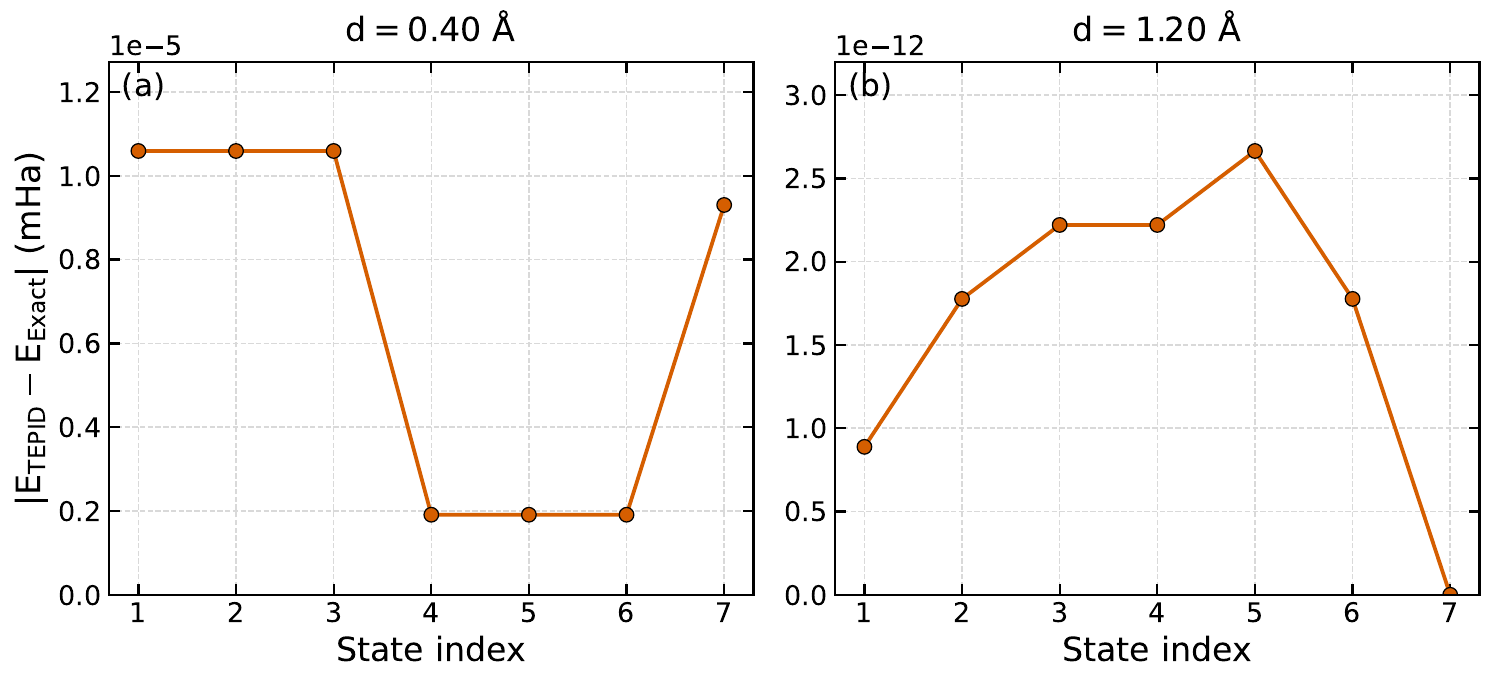}
\caption{Absolute energy errors for the energy eigenstates of H$_4$ corresponding to Fig.~\ref{fig:energy_level_structure2}. Errors are shown in mHa as a function of state index and the graphs show consistently small deviations across states from the exact results within chemical accuracy at (a) $d=0.40$~\AA{} and (b) $d=1.20$~\AA{}.}
\label{fig:H4_error}
\end{figure*}

Similarly, Fig.~\ref{fig:H4_error} presents the absolute energy errors for the energy states of H$_4$, corresponding to Fig.~\ref{fig:energy_level_structure2}. As in the H$_2$ case, the errors are shown in mHa as a function of the state index. The results again exhibit small deviations across all states, indicating strong agreement with the exact energies. Notably, chemical accuracy is maintained at both $d=0.40$~\AA{} and $d=1.20$~\AA{} again by the TEPID-ADAPT algorithm.

\

\

\

\end{appendix}

\end{document}